\documentclass{article}
\usepackage{graphics}
 \usepackage{graphicx}

 \usepackage{epsfig}

\usepackage{amssymb}
\usepackage{amsmath}
 \usepackage{amsthm}

\begin{document}

\title{Spin-current rectification through a quantum dot using temperature bias}

\author{M.˜ Bagheri Tagani,
        H.˜ Rahimpour Soleimani,\\
        \small{Department of physics, University of Guilan, P.O.Box 41335-1914, Rasht, Iran}}

\maketitle

\begin{abstract}
We analyze spin-dependent transport through a spin-diode in the
presence of spin-flip and under influence of temperature bias.
The current polarization and the spin accumulation are
investigated in detail by means of reduced density matrix.
Results show that the spin accumulation is linearly increased
when the metallic electrode is warmer whereas, its behavior is
more complicated when the ferromagnetic lead is warmer.
Furthermore, spin-flip causes that the current polarization
becomes not only a function of spin-flip rate but also a function
of temperature. The current polarization is reduced up to 90\% if
the time of spin-flip  is equal to the tunneling time. The
behavior of spin-dependent current is also studied as a function
of temperature, spin-flip rate, and polarization.
\end{abstract}

\section{Introduction}
\label{Introduction} Study of electron transport through devices
fabricated from quantum dots (QDs) has attracted a lot of
attention during two recent
decades~\cite{Ref1,Ref2,Ref3,Ref4,Ref5,Ref6,Ref7,Ref8,Ref9}.
Transport through QDs exhibits novel and interesting phenomena
such as Coulomb and spin blockade
effects~\cite{Ref10,Ref11,Ref12,Ref13}, Kondo
effect~\cite{Ref14,Ref15,Ref16}, negative differential
conductance~\cite{Ref17,Ref18}, and so on. Coupling of QD to
different leads such as, normal metal, ferromagnet, or
superconductor, is now feasible due to recent progress in
nanotechnology. One of the most interesting configurations is a
quantum dot coupled to a normal metal(NM) and a ferromagnetic
(FM) lead. In recent years the configuration has been extensively
studied experimentally and
theoretically~\cite{Ref19,Ref20,Ref21,Ref22,My1}. With regard to
spin-current rectification effect observed in the device, it can
work as a spin-diode.
\par the most studies done about the spin-diode have been focused on
the behavior of the system in the presence of electric bias. Due
to recent advance in the field of thermoelectricity, it is now
possible to produce the spin current by applying temperature bias
across a ferromagnetic semiconductor~\cite{Ref23,Ref24} or a
magnetic insulator~\cite{Ref25}. Very recently, F. Qi and
co-workers~\cite{Ref26} have studied the transport through a QD
coupled to a normal metal and a ferromagnetic electrode under
influence of the temperature bias. they have reported that the
temperature gradient results in a rectification effect in the
current polarization. In this article, we analyze the behavior of
the system in the presence of spin-flip. Transverse magnetic
field~\cite{Ref27}, spin-orbit interaction~\cite{Ref28} and etc
can cause the spin-flip. With regard to the structure of the
device, it is completely possible that the electron of the QD
interacts with the polarized nucleus of the sublayer. Current
polarization, spin accumulation, and spin current are analyzed in
detail by means of reduced density matrix
approach~\cite{Ref29,Ref30}. It is found that spin-flip affects
current polarization significantly.
\par In the next section, model Hamiltonian and equations used for
calculations are presented. Using density matrix approach and
wide band limit, formal expressions of electron density and
current are given. Numerical results and analytical expressions
for the spin accumulation and the current polarization are
presented in section 3. In the end, some sentences are given as
conclusion.
\begin{figure}[htb]
\begin{center}
\includegraphics[height=7cm,width=9cm,angle=0]{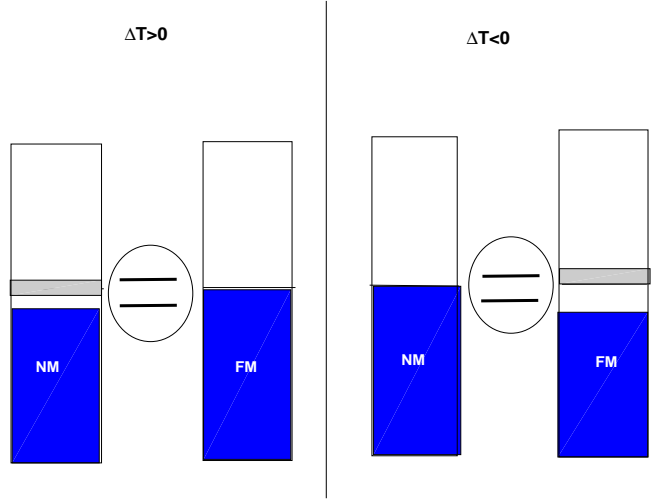}\label{fig:1}       
\caption{Schematic diagram of a spin diode. $\Delta{T}$ denotes
the temperature difference between leads. The left lead is warmer
if $\Delta{T}>0$ and $\Delta{T}<0$ means the right lead is warmer.
The increase of temperature causes that the electrons of the lead
occupy the states above the chemical potential shown as gray
space.}
\end{center}
\end{figure}

 \section{Model}
 \label{Model}
 We consider a single level quantum dot coupled to a normal metal
 (NM) and a ferromagnetic (FM) lead as fig. 1. The Hamiltonian
 describing the system is given as follows
\begin{align}\label{Eq.1}
  H&=\sum_{\alpha k\sigma}\varepsilon_{\alpha k\sigma}c^{\dag}_{\alpha
  k\sigma}c_{\alpha k\sigma}+\sum_{\sigma} \varepsilon_{\sigma}
  n_{\sigma}+Un_{\uparrow}n_{\downarrow}+\\ \nonumber
  &
  R[d^{\dag}_{\uparrow}d_{\downarrow}+d^{\dag}_{\downarrow}d_{\uparrow}]+\sum_{\alpha
  k\sigma}[V_{\alpha k\sigma}c^{\dag}_{\alpha k\sigma}d_{\sigma}+H.C]
\end{align}
where non-interacting quasi-particle approximation is used to
describe the electrodes, $c_{\alpha k\sigma}(c^{\dag}_{\alpha
k\sigma})$ destroys (creates) an electron with wave vector $k$,
spin $\sigma$, and energy $\varepsilon_{\alpha k\sigma}$ in the
lead $\alpha$ $(\alpha=L,R)$. $d_{\sigma}(d^{\dag}_{\sigma})$ is
annihilation (creation) operator in the QD which destroys
(creates) an electron with spin $\sigma$ in the QD. $U$ denotes
on-site Coulomb repulsion and
$n_{\sigma}=d^{\dag}_{\sigma}d_{\sigma}$ is occupation operator.
The fourth term describes spin-flip process and $R$ is spin-flip
rate which is spin-independent. The last term describes tunneling
between the QD and the electrodes and $V_{\alpha k\sigma}$ stands
for the coupling strength.
\par It is clear that the QD can be in : empty state $|0>$, singly
occupied state $|\sigma>$ ($\varepsilon_\sigma$), or doubly
occupied state $|2>$
($\varepsilon_2=\varepsilon_\uparrow+\varepsilon_\downarrow+U$).
States $|\uparrow>$ and $|\downarrow>$ are not the eigenstates of
the isolated QD Hamiltonian because of spin-flip. Using Markov
approximation the time evolution of the density matrix elements
are given as~\cite{Ref29,Ref30}
\begin{align}\label{Eq.2}
  \frac{dP_{ss'}}{dt}&=-i<s|[H_{s.f},P]|s'>+\delta_{ss'}\sum_{k\neq
  s}[W_{ks}P_{kk}-W_{sk}P_{ss}]\\ \nonumber
  & -\frac{1}{2} [1-\delta_{ss'}][\sum_{k \neq s}W_{sk}+\sum_{s'\neq k}W_{s'
  k}]P_{ss'}
\end{align}
where $H_{s.f}$ denotes the fourth term in Eq.\eqref{Eq.1}.
$P_{ss}$ is probability of being in the state $s$ ($s=0,\sigma,2$)
whereas, $P_{ss'}$ ($s \neq s'$) describes coherency between
states $|\uparrow>$ and $|\downarrow>$. $W_{ss'}$ stands for
transition from state $|s>$ to $|s'>$ and is computed by Fermi's
golden rule as
\begin{equation}\label{Eq.3}
  W_{ss'}=\sum_{\alpha}\Gamma_{\sigma}^{\alpha}[f_{\alpha}(|\varepsilon_{ss'}|)\delta_{N_s,N_{s'}+1}+
  \bar{f}_{\alpha}(|\varepsilon_{ss'}|)\delta_{N_s,N_s'-1}]
\end{equation}
where $N_s$ denotes the number of electrons in the state $s$ and
$\varepsilon_{ss'}=\varepsilon_s-\varepsilon_s'$.
$f_{\alpha}(x)=(1+exp((x-\mu_\alpha)/k_BT_\alpha))^{-1}$ is Fermi
distribution function of lead $\alpha$ where $T_\alpha$ and
$\mu_\alpha$ are temperature and chemical potential of lead
$\alpha$ and $\bar{f_{\alpha}}=1-f_\alpha$.
$\Gamma^\alpha_\sigma$ is spin-dependent tunneling rate of the
lead $\alpha$  obtained by using wide band approximation. As an
instance, $P_{{\uparrow}{\downarrow}}$ is obtained from
Eq.\eqref{Eq.2} as
\begin{equation}\label{new}
 \frac{dP_{{\uparrow}{\downarrow}}}{dt}=iR[P_{\uparrow\uparrow}-P_{\downarrow\downarrow}]-
 \frac{1}{2}[\sum_{k \neq \uparrow}W_{\uparrow k}+\sum_{k \neq \downarrow}W_{\downarrow
 k}]P_{{\uparrow\downarrow}}
\end{equation}
and $P_{\downarrow\uparrow}=P_{\uparrow\downarrow}^*$.
\par Now, we define $n_{\sigma}=P_{\sigma\sigma}+P_{22}$. Using
Eq.\eqref{Eq.2} and normalization condition
($P_{00}+\sum_{\sigma}P_{\sigma\sigma}+P_{22}=1$), it is
straightforward to show
\begin{equation}\label{Eq.4}
  \frac{dn_{\sigma}}{dt}=W_{0\sigma}[1-n_{\sigma}-n_{\bar{\sigma}}]-W_{\sigma
  0}n_{\sigma}+W_{\bar{\sigma}2}n_{\bar{\sigma}}+iR[P_{\sigma\bar{{\sigma}}}-P_{\bar{\sigma}\sigma}]
\end{equation}
where $\bar{\sigma}$ is opposite of $\sigma$. Solving
Eq.\eqref{Eq.4} in the steady state ($\frac{dn_{\sigma}}{dt}=0$),
the spin-dependent current crossing from the lead $\alpha$ is
given as
\begin{equation}\label{Eq.5}
  I^{\sigma}_{\alpha}=W^{\alpha}_{0\sigma}[1-n_{\sigma}-n_{\bar{\sigma}}]-W^{\alpha}_{\sigma
  0}n_{\sigma}+W^{\alpha}_{\bar{\sigma}2}n_{\bar{\sigma}}
\end{equation}
\par In the continue, we consider $\Gamma_0=10\mu{eV}$ as lead-QD tunneling
rate and set $\Gamma^{L}_{\sigma}=\Gamma_0$ and
$\Gamma^{R}_{\sigma}=\Gamma_0[1+\sigma \rho]$ where
$\sigma=1(-1)$ for $\uparrow(\downarrow)$, and $\rho$ is the spin
polarization degree of the ferromagnetic lead. Note that the left
lead is a normal meta thus its tunneling rate is spin-independent
whereas, the right lead is a ferromagnet whose majority carriers
are assumed to be spin-up electrons. In addition, we assume
$\varepsilon_\sigma=-2meV$, $U=5meV$, and $\Delta{T}=T_L-T_R$.
$\Delta{T}>0$ means the left lead is warmer ($T_L=T_0+\Delta{T}$)
while, $\Delta{T}<0$ means $T_R=T_0+\Delta{T}$ where $T_0=0.2
meV$ is the base temperature of the system.

\begin{figure}[htb]
\begin{center}
\includegraphics[height=8cm,width=8cm,angle=0]{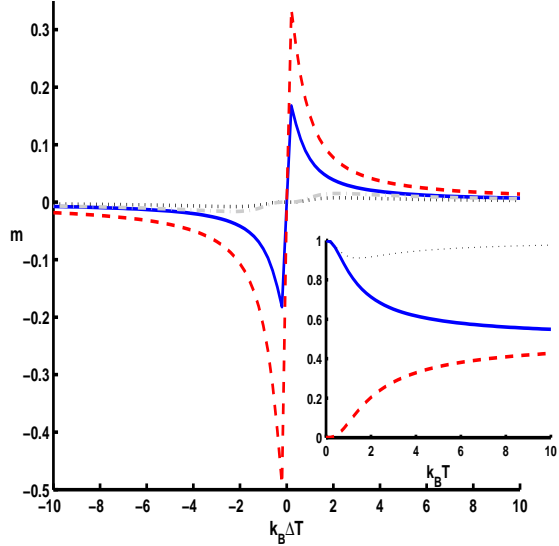}\label{fig:2}       
\caption{Spin accumulation versus temperature bias. $S=0$ and
$\rho=0.4$ (solid), $S=0$ and $\rho=0.8$ (dashed), $S=0.5$ and
$\rho=0.4$ (dotted) and $S=0.5$ and $\rho=0.8$ (dash-dotted).
Inset shows $\alpha$ (solid), $\beta$ (dashed), and
$\alpha+\beta$ (dotted).}
\end{center}
\end{figure}

\section{Results and discussions}
\label{Numerical results} In order to analyze the spin
accumulation and the current, we have assumed that
$f_{R}(\varepsilon_\sigma)=1$, $f_R(\varepsilon_\sigma+U)=0$,
$f_{L}(\varepsilon_\sigma)=\alpha$ and
$f_{L}(\varepsilon_\alpha+U)=\beta$ for $\Delta{T}>0$ whereas,
$f_{L}(\varepsilon_\sigma)=1$, $f_L(\varepsilon_\sigma+U)=0$,
$f_{R}(\varepsilon_\sigma)=\alpha$ and
$f_{R}(\varepsilon_\alpha+U)=\beta$ in $\Delta{T}<0$. Note that
$\alpha$ and $\beta$ are temperature-dependent and their
dependence on the temperature is shown in inset of fig. 2. Fig. 2
shows the spin accumulation ($m=n_{\uparrow}-n_{\downarrow}$) as a
function of temperature bias. Results show that the most change
in $m$ is happened when the temperature difference is very low.
As one can see in inset of fig. 2, $\alpha\rightarrow 1$ and
$\beta \rightarrow 0$ in low temperature, so that there is only
one electron in the QD and as a result, the ferromagnetic
electrode affects the spin properties of the system
significantly. By increase of temperature, the second electron
enters the QD thus the spin characteristics of the system are
vanished. By solving Eq.\eqref{Eq.4} in the steady state we
obtain blow equation for the spin accumulation as a function of
temperature, polarization, and spin-flip rate.
\begin{subequations}\label{new}
\begin{align}
m&=\frac{2\rho[(\alpha-1)^2-\beta^2]}{4S^2[3+\alpha-\beta]+y[4-(\alpha+1-\beta)^2]}
\quad \Delta{T}>0\\
m&=\frac{2\rho[\alpha+\beta-1]y}{\rho^2y[(\alpha-\beta)^2-1]+4S^2[3+\alpha-\beta]-y[(\alpha-\beta+1)^2-4]}
\quad \Delta{T}<0
\end{align}
\end{subequations}
where $y=1+\beta-\alpha$.
\par From Eq.\eqref{new}, it is obvious that increase of
temperature difference results in $m=0$ because of
$\alpha,\beta\rightarrow 1/2$. Physically, by warming an
electrode, electrons near the Fermi surface occupy states above
the chemical potential, see fig. 1, so that there will be some
holes below the chemical potential. The excited electrons have
the necessary energy to overcome the charging energy so that the
QD will have two electrons and as a consequence, $m=0$.
Eq.\eqref{new} also shows that the dependence of $m$ on $\rho$
and $S$ is different in positive and negative temperature bias
and this difference results in an asymmetry in spin
accumulation-temperature bias characteristic of the system. As
one can see in Eq.\eqref{new}, $m$ decreases rapidly by increase
of $S$ hence the variations of the spin accumulation become
inconsequential in the presence of spin-flip, as it is shown in
fig. 2. In positive and low temperature difference, $m$ is
positive because the ferromagnetic lead acts as an emitter and
due to $\Gamma_R^\uparrow>\Gamma_R^\downarrow$, the probability
of being in the spin-up state is more. Note that
$\alpha+\beta<1$, see dotted line in the inset, and as a result
$(\alpha-1)^2-\beta^2$ becomes positive in Eq.\eqref{new}. In
negative and low temperature difference, the left lead acts as
the emitter and as a result, $m$ becomes negative because the
spin-down electron has to stay in the QD for a longer time.
\begin{figure}[htb]
\begin{center}
\includegraphics[height=7cm,width=9cm,angle=0]{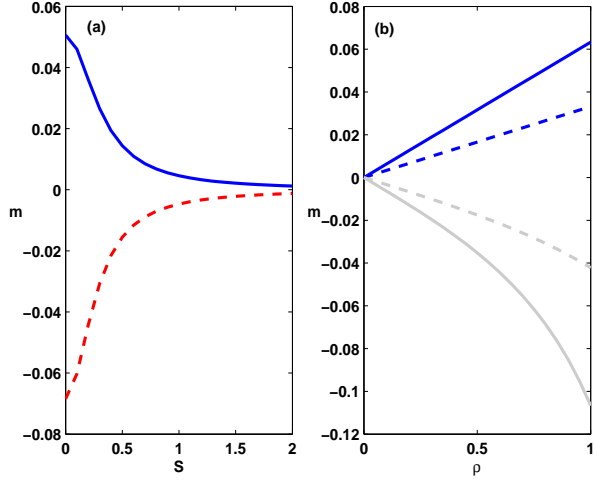}\label{fig:3}       
\caption{(a) $m$ against $S$ for $\Delta{T}=3$ (solid) and
$\Delta{T}=-3$(dashed). (b) $m$ versus $\rho$ for $S=0$ (solid)
and $S=0.3$ (dashed). $\Delta{T}=-3$ is shown in gray.}
\end{center}
\end{figure}
\par The dependence of the spin accumulation on the spin-flip rate
and the polarization is plotted in fig. 3. asymmetry of $m$
versus $R$ is well seen in fig. 3a. As expect from
Eq.\eqref{new}, $m$ is significantly reduced by increase of $R$.
The dependence of $m$ on $\rho$ is much more interesting. In
positive temperature bias, the spin accumulation is linearly
increased by increase of $\rho$ whereas, its behavior is
completely different in negative temperature difference. The main
result is risen due to the change of the emitter. In
$\Delta{T}>0$ where the right lead is emitter, the linear
relation is dominant while, in the condition that the left lead
is emitter, the dependence is more complicated. We reported the
same behavior for the spin accumulation in a spin-diode under
influence of voltage bias~\cite{My2}.
\begin{figure}[htb]
\begin{center}
\includegraphics[height=9cm,width=9cm,angle=0]{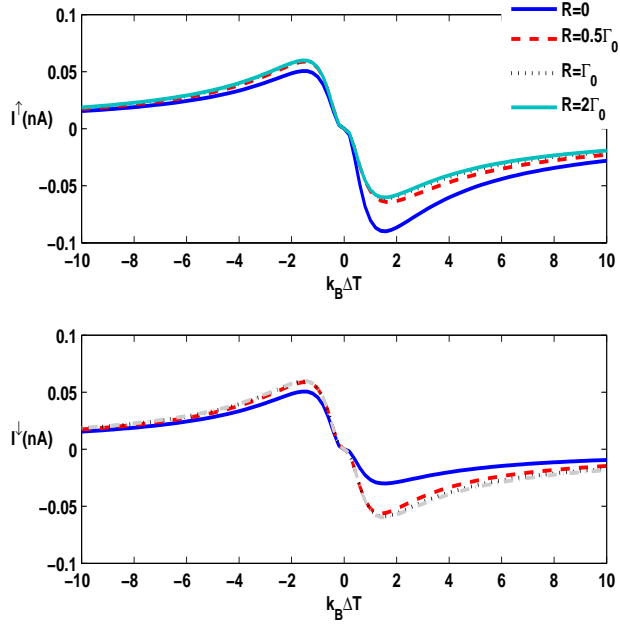}\label{fig:4}       
\caption{Spin-resolved currents versus temperature bias.}
\end{center}
\end{figure}
\par Spin-resolved currents as a function of temperature bias,
spin-flip rate, and polarization are shown in figs. 4 and 5,
respectively. First, we estimate the current in the presence of
positive temperature bias. By solving Eq.\eqref{Eq.4} and with
respect to Eq.\eqref{Eq.5}, spin-dependent currents are obtained
as:
\begin{subequations}\label{current}
\begin{align}
I_{L}^{\uparrow}&=\Gamma_0[\alpha-n_{\uparrow}-(\alpha-\beta)n_{\downarrow}]
\\
I_{L}^{\downarrow}&=\Gamma_0[\alpha-(\alpha-\beta)n_{\uparrow}-n_{\downarrow}]
\end{align}
\end{subequations}
where $n_{\sigma}$ is given as
\begin{equation}\label{density}
  n_{\sigma}=\frac{\rho\sigma[(1-\alpha)^2-\beta^2]+4S^2(1+\alpha)+(\beta+1)^2+\alpha[(\alpha-\beta)^2-(1+\alpha)]}
  {4S^2[3+\alpha-\beta]+y[4-(1+\alpha-\beta)^2]}
\end{equation}
where $\sigma=1(-1)$ for $\uparrow(\downarrow)$. In positive
bias, the current crossing from the left lead is negative because
the right lead acts as the emitter, as explained before. From
Eq.\eqref{density}, it is clear that $n_{\uparrow}$ is reduced by
increase of $R$ whereas, increase of $R$ gives rise to increasing
$n_{\downarrow}$. This effect results in reduction of
$I^{\uparrow}$ and increase of $I_{\downarrow}$, as it is seen in
Eq.\eqref{current}. Note that the probability of finding the QD
in spin-up state is more in positive temperature bias if spin-flip
process is absent. In the presence of spin-flip, the spin-up
electron may rotate and change to a spin-down electron.
Therefore, the spin-flip increases the probability of being in
the spin-down while, the probability of being in spin-up state is
reduced. With respect to Eq.\eqref{density}, although the
existence of $R$ influences significantly on the current, the
magnitude of it does not have significant role in the curvature of
the current due to existence of a large term in denominator.
Furthermore, $I^{\uparrow}$ will be equal to $I^{\downarrow}$ if
$R>>\Gamma_0$ because of $n_{\uparrow}=n_{\downarrow}$, see
Eq.\eqref{density}. From Eqs.(8,9), it is obvious that the current
is related to $\rho$ linearly. This dependence is well observed in
fig. 5b. It is very interesting to note that unlike $R$, increase
of $\rho$ results in increase of $I^{\uparrow}$ and decrease of
$I^{\downarrow}$. Indeed, with increase of $\rho$ the spin-up
electron is injected into the QD faster and as a result
$I^{\uparrow}>I^{\downarrow}$.
\begin{figure}[htb]
\begin{center}
\includegraphics[height=9cm,width=10cm,angle=0]{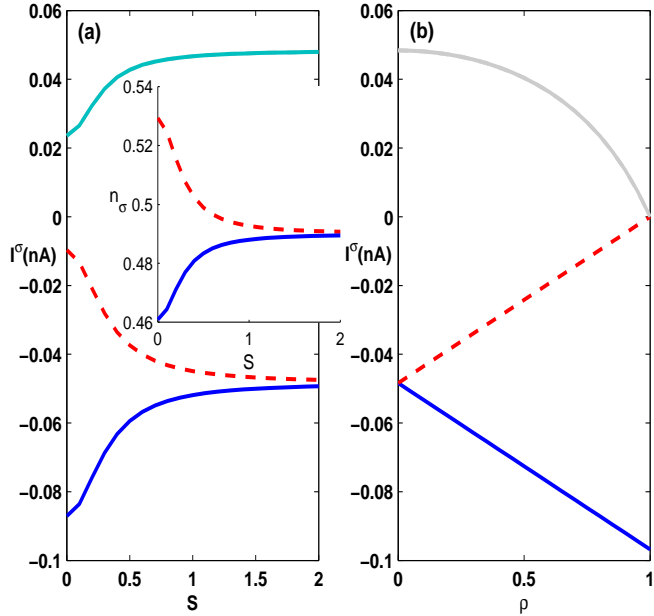}\label{fig:5}       
\caption{(a) $I^{\sigma}$ as a function of spin-flip rate. We set
$\rho=0.8$ and $\Delta{T}=\pm 3$. Solid line is $I^{\uparrow}$
and dashed line is $I^{\downarrow}$. $\Delta{T}=-3$ is plotted in
gray. (b) spin-dependent current versus $\rho$. We set $R=0$.
Inset shows $n_{\uparrow}$ (solid) and $n_{\downarrow}$ (dashed)
for $\Delta{T}=-3$.}
\end{center}
\end{figure}
\par Now, we analyze the current when the ferromagnetic lead is
warmer. In this case, $I^{\uparrow}$ is equal to $I^{\downarrow}$
and given as
\begin{equation}\label{current2}
  I^{\sigma}=\Gamma_0[1-n_{\uparrow}-n_{\downarrow}]
\end{equation}
where $n_{\sigma}$ is obtained from
\begin{equation}\label{density2}
  n_{\sigma}=\frac{\rho\sigma y[\alpha+\beta-1]-\alpha
  y^2\rho^2+4S^2[1+\alpha]+y^2[1+\alpha]}{\rho^2 y[(\alpha-\beta)^2-1]+4S^2(3+\alpha-\beta)-y[(1+\alpha-\beta)^2-4]}
\end{equation}
The first result observed in fig. 4 is the current is positive
i.e. the left lead acts as the emitter. It comes from the fact
that electrons below the Fermi level in the right lead are lesser
than electrons in the left lead. On the other hand, increase of
$R$ results in increase of both $I^{\uparrow}$ and
$I^{\downarrow}$. From inset of fig. 4, it is clear that increase
of $R$ gives rise to reduction of $n_{\downarrow}$ and increase
of $n_{\uparrow}$. Note that the probability of being in the
spin-down state is more if $R=0$ because the spin-up electron
injected from the left lead leaves the QD faster than the
spin-down electron. As it is obvious in inset and
Eq.\eqref{density2}, the rate of reducing $n_{\downarrow}$ is
faster than the rate of increase of $n_{\uparrow}$. This fact
leads to increase of $I^{\sigma}$ in the presence of spin-flip.
Increase of $I^{\sigma}$ due to spin-flip is clearly seen in fig.
5a. Like $\Delta{T}>0$, the spin-resolved currents saturate for
$R>0.5\Gamma_0$. Unlike $\Delta{T}>0$, the dependence of
$I^{\sigma}$ on the polarization is not linear anymore. This
dependence is plotted in fig. 5b. The behavior of $I^{\sigma}$ is
very interesting in $\rho=1$. In positive temperature bias,
$I^{\downarrow}$ becomes zero because there are no spin-down
electrons in ferromagnetic lead to enter the QD, but
$I^{\uparrow}=I^{\downarrow}=0$ if $\Delta{T}<0$. It is
straightforward to show that $n_{\uparrow}+n_{\downarrow}=1$ if
$S=0$ and $\rho=1$ and with respect to Eq.\eqref{current2},
$I^{\sigma}$ becomes zero. Indeed, the electron inside the QD
cannot tunnel to the right lead so,  other electron cannot enter
the QD. In addition, increase of $\rho$ results in decrease of
$I^{\sigma}$.
\begin{figure}[htb]
\begin{center}
\includegraphics[height=10cm,width=10cm,angle=0]{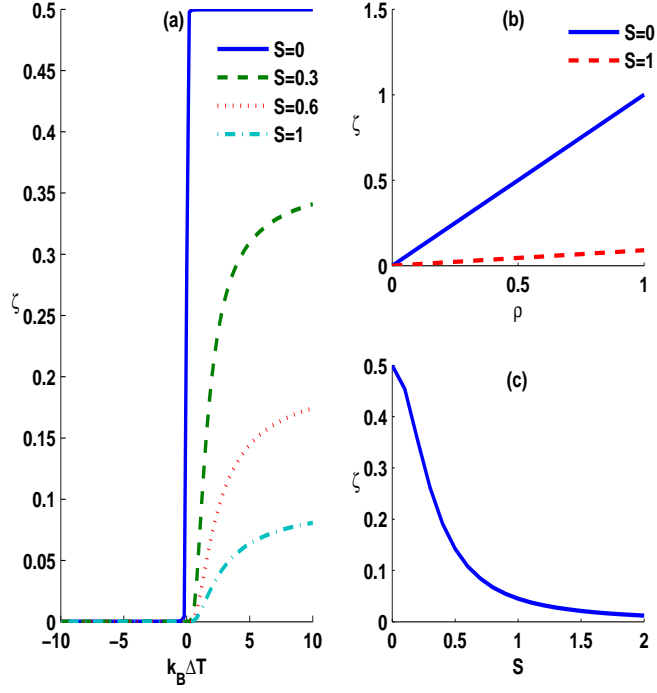}\label{fig:6}       
\caption{Current polarization versus (a) temperature bias, (b)
polarization, and (c) spin-flip rate.}
\end{center}
\end{figure}
\par The current polarization,
$\zeta=\frac{I^{\uparrow}-I^{\downarrow}}{I^{\uparrow}+I^{\downarrow}}$
,is plotted in fig. 6a. With respect to Eqs.(8,10), it is
straightforward to show that
\begin{subequations}\label{polarization}
\begin{align}
  \zeta&=\frac{y^2}{y^2+4S^2}\rho \quad for \quad \Delta{T}>0\\
  \zeta&=0 \qquad for \quad \Delta{T}<0
\end{align}
\end{subequations}
Note that our results are identical results given in
Ref.~\cite{Ref26} if $S=0$ and $\alpha=\beta=1/2$. Presence of
spin-flip causes that the current polarization becomes not only a
function of spin-flip rate but also a function of temperature
because of presence of $\alpha$ and $\beta$ in
Eq.\eqref{polarization}. As expected, spin-flip gives rise to
reduction of $\zeta$ because this process tries to destroy the
spin characteristics of the system. If spin-flip rate becomes
equal to tunneling rate, $\zeta$ is reduced up to $90\%$.
However, the system can still work as a rectifier although its
performance is weak. The dependence of $\zeta$ on polarization
and $R$ is shown in figs. 6b and 6c, respectively. Linear
behavior of $\zeta$ versus $\rho$ is well seen. $\zeta=1$ if
$R=0$ and $\rho=1$ because there is no spin-down current through
the system but, $\zeta$ is significantly reduced in the presence
of spin-flip because the spin-up electron into the QD can change
to a spin-down electron and as a result, the spin-down current is
created. Unlike fig. 6b, the behavior of $\zeta$ against $R$ is
nonlinear. Such behavior about current polarization was before
reported in the presence of electric bias~\cite{My2}.

\section{Conclusion}
\label{conclusion} In this article, We study spin-dependent
transport through a quantum dot coupled to a normal metal and a
ferromagnetic electrode. Analytical expressions for spin
accumulation and current polarization are obtained in the
presence of spin-flip, using master equations. Results show that
the dependence of the system on the polarization is linear when
the metallic lead is warmer whereas, its behavior is more
complicated when ferromagnet is warmer. On the other hand, the
spin-flip affects significantly the current polarization so that
it is reduced up to 90\% if the spin-flip time is equal to
tunneling time. The effects of temperature, polarization, and
spin-flip rate on the spin-resolved currents are also estimated.
When the metallic electrode is warmer, spin-flip results in the
increase of spin-down current while, spin-up current is decreased.

\bibliographystyle{model1a-num-names}
\bibliography{<your-bib-database>}

\begin{thebibliography}{00}

\bibitem{Ref1}
W.G. van der Wiel, S. De Franceschi, J.M. Elzerman, T. Fujisawa,
S. Tarucha, L.P. Kouwenhoven, Electron transport through double
quantum dots, Rev. Mod. Phys. 75 (2002) 1.

\bibitem{Ref2}
R. Hanson, L.P. Kouwenhoven, J.R. Petta, S. Tarucha, L.M.K.
Vandersypen, Spins in few-electron quantum dots,  Rev. Mod. Phys.
79 (2007) 1217.

\bibitem{Ref3}
Y. Meir, N. S. Wingreen, P. A. Lee, Low-temperature transport
through a quantum dot: The Anderson model out of equilibrium,
Phys. Rev. Lett. 70 (1993) 2601.

\bibitem{Ref4}
K. Ono, D.G. Austing, Y. Tokura, S. Tarucha, Current
Rectification by Pauli Exclusion in a Weakly Coupled Double
Quantum Dot System,  Science 297 (2002) 1313.

\bibitem{Ref5}
J. M. Elzerman, R. Hanson, L. H. W. van Beveren, B. Witkamp, L.
M. K. Vandersypen, L. P. Kouwenhoven, Single-shot read-out of an
individual electron spin in a quantum dot, Nature 430 (2004) 431.

\bibitem{Ref6}
A. C. Johnson, J. R. Petta, J. M. Taylor, A. Yacoby, M. D. Lukin,
C. M. Marcus, M. P. Hanson,  A. C. Gossard, Triplet-singlet spin
relaxation via nuclei in a double quantum dot,  Nature 435 (2005)
925.

\bibitem{Ref7}
M.H. Mikkelsen, J. Berezovsky, N.G. Stoltz, L.A. Coldren, D.D.
Awschalom, Optically detected coherent spin dynamics of a single
electron in a quantum dot, Nature Phys. 3 (2007) 770.

\bibitem{Ref8}
N. Shaji, C. B. Simmons, M. Thalakulam, L. J. Klein, H. Qin, H.
Luo, D. E. Savage, M. G. Lagally, A. J. Rimberg, R. Joynt, M.
Friesen, R. H. Blick, S. N. Coppersmith, M. A. Eriksson, Spin
blockade and lifetime-enhanced transport in a few-electron
Si/SiGe double quantum dot, Nature Phys. 4 (2008) 540.

\bibitem{Ref9}
H.Z. Lu, S.Q. Shen, Using spin bias to manipulate and measure
spin in quantum dots, Phys. Rev. B 77 (2008) 235309.

\bibitem{Ref10}
H.W. Liu, T. Fujisawa, Y. Ono, H. Inokawa, A. Fujiwara, K.
Takashina, Y. Hirayama, Pauli-spin-blockade transport through a
silicon double quantum dot, Phys. Rev. B 77 (2008) 073310.

\bibitem{Ref11}
C.Romeike, M.R. Wegewijs, M. Ruben, W. Wenzel, H. Schoeller,
Charge-switchable molecular magnet and spin blockade of tunneling,
Phys. Rev. B 75 (2007) 064404.

\bibitem{Ref12}
  J. Park, A.N. Pasupathy, J.I.
Goldsmith, C. Chang, Y. Yaish, J.R. Petta, M. Rinkoski, J.P.
Sethna, H.D. Abruña, P.L. McEuen, D.C. Ralph, Coulomb blockade
and the Kondo effect in single-atom transistors, Nature 417 (2002)
722.

\bibitem{Ref13}
 B. Dong, H.L. Cui, X.L. Lei, Quantum rate equations for electron
 transport through an interacting system in the sequential tunneling regime, Phys. Rev. B 69 (2004) 035324.

\bibitem{Ref14}
  W. Liang, M.P. Shores, M. Bockrath, J.R. Long, H. Park, Kondo resonance in a single-molecule transistor, Nature 417 (2002)
725.

\bibitem{Ref15}
 Takahide Numata, Yunori Nisikawa, Akira Oguri, Alex C. Hewson, Kondo effects in a triangular triple quantum dot: Numerical renormalization group study in the whole region of the electron
 filling,
Phys. Rev. B 80 (2009) 155330.

\bibitem{Ref16}
 I. Weymann, J. Barna\'{s}, Kondo effect in a quantum dot coupled to ferromagnetic leads and side-coupled to a nonmagnetic reservoir Phys. Rev. B 81 (2010) 035331.

 \bibitem{Ref17}
J.N. Pedersen, B. Lassen, A. Wacker, Coherent transport through
an interacting double quantum dot: Beyond sequential tunneling,
Phys. Rev. B 75 (2007) 235314.

\bibitem{Ref18}
F. Elste, D. R. Reichman,  A. J. Millis, Transport through a
quantum dot with two parallel Luttinger liquid leads, Phys. Rev. B
83 (2011) 245405.

\bibitem{Ref19}
C.A. Merchant, N. Markoviæ\'{c}, Current and Shot Noise
Measurements in a Carbon Nanotube-Based Spin Diode, J. Appl.
Phys. 105 (2009) 07c711.

\bibitem{Ref20}
A. Iovan, S. Andersson, Yu.G. Naidyuk, A. Vedyaev, B. Dieny, V.
Korenivski, Spin Diode based on Fe/MgO Double Tunnel Junction,
Nano Lett. 8 (3) (2008) 805.

\bibitem{Ref21}
F.M. Souza, J.C. Egues, A.P. Jauho, Quantum dot as a spin-current
diode: A master-equation approach, Phys. Rev. B 75 (2007) 165303.

\bibitem{Ref22}
I. Weymann, J. Barnas, Spin diode based on a single-wall carbon
nanotube, J. Appl. Phys. Lett. 92 (2008) 103127.

\bibitem{My1}
M. B. Tagani, H. R. Soleimani, Study of coupling effect on the
performance of a spin-current diode: Nonequilibrium Green's
function based model, Solid State Commun. 151 (2011) 1479.

\bibitem{Ref23}
K. Uchida, S. Takahashi, K. Harii, J. Ieda, W. Koshibae, K. Ando,
S. Maekawa, and E. Saitoh, Observation of the spin Seebeck effect
, Nature  455 (2008) 778.

\bibitem{Ref24}
C. M. Jaworski, J. Yang, S. Mack, D. D. Awschalom, J. P.
Heremans, R. C. Myers, Observation of the spin-Seebeck effect in
a ferromagnetic semiconductor, Nat. Mater 9 (2010) 898.

\bibitem{Ref25}
K. Uchida, J. Xiao, H. Adachi, J. Ohe, S. Takahashi, J. Ieda, T.
Ota, Y. Kajiwara, H. Umezawa, H. Kawai, G. E. W. Bauer, S.
Maekawa, E. Saitoh, Spin Seebeck insulator , Nat. Mater 9 (2010)
894.

\bibitem{Ref26}
F. Qi, Y. Ying, G. Jin, Temperature-manipulated spin transport
through a quantum dot transistor, Phys. Rev. B 83 (2011) 75310.

\bibitem{Ref27}
H. A. Engel and D. Loss, Single-spin dynamics and decoherence in
a quantum dot via charge transport, Phys. Rev. B 65 (2002) 195321.

\bibitem{Ref28}
J. Danon and Yu. V. Nazarov,Pauli spin blockade in the presence
of strong spin-orbit coupling, Phys. Rev. B 80 (2009) 41301.

\bibitem{Ref29}
K. Blum, Density Matrix Theory and Applications (Plenum, New
York, 1981).

\bibitem{Ref30}
G. Mahler and V. A. Weberru\ss, Quantum Networks: Dynamics of Open
Nanostructures (Springer, Berlin, 1995).

\bibitem{My2}
M. B. Tagani, H. R. Soleimani, Influence of spin-flip on the
performance of the spin-diode, MPLB. 25 (2011) 2335.


 \end{thebibliography}

\end{document}